\documentclass[preprintnumbers,amsmath,amssymb,groupedaddress,floatfix]{revtex4}

\usepackage{graphicx}
\usepackage{bm}
\usepackage{slashed}
\usepackage{amsfonts}
\usepackage{graphicx,color}
\usepackage[colorlinks,linkcolor=blue,citecolor=green]{hyperref}
\usepackage{wrapfig}
\newcommand{\dd}{\textrm{\,d}}
\newcommand{\ee}{\textrm{\,e}}

\makeatletter
\@addtoreset{figure}{section}
\makeatother

\begin{document}

\title{Hadron Spectra Parameters within the Non-Extensive~Approach}

\author{Keming Shen}
\email{shen.keming@wigner.mta.hu}

\author{Gergely G\'abor Barnaf\"oldi}

\author{Tam\'as S\'andor Bir\'o}
\affiliation{Wigner Research Center for Physics of the HAS, 29-33 Konkoly-Thege Mikl\'os Street, 1121 Budapest, Hungary}

\date{\today}

\begin{abstract}

We investigate how the non-extensive approach works in high-energy physics. Transverse momentum ($p_T$) spectra of several hadrons are fitted by various non-extensive momentum distributions and by the Boltzmann--Gibbs statistics.~It is shown that some non-extensive distributions can be transferred one into another.~We find explicit hadron mass and center-of-mass energy {scaling both in the temperature and in the non-extensive parameter, $q$,} in proton--proton and heavy-ion collisions. 
We find that the temperature depends linearly, but the Tsallis $q$ follows a logarithmic dependence on the collision energy in proton--proton collisions.
In the nucleus--nucleus collisions, on~the other hand, $T$ and $q$ correlate linearly, as was predicted in our previous work.

\end{abstract}

\maketitle

\section{Introduction}\label{sec:sec1}

In high-energy nuclear physics, the investigation of transverse momentum ($p_T$) spectra is a fundamental measure in statistical approaches. 
The $p_T$ spectrum reveals information on the kinetic properties of the particles produced in high-energy collisions.
Strong correlation phenomena were recently observed in proton--proton and heavy-ion collisions~\cite{corr-1,corr-2}, their statistical and thermodynamical description points beyond the classical Boltzmann--Gibbs (BG) statistics.
It has  long been realized that data on single inclusive particle distributions show a power-law behavior in the high-$p_T$ region.
For~these, the Pareto--Hagedorn--Tsallis distribution has been frequently applied~\cite{app-p0,app-p00,app-p1}.~Its form coincides with the generalized $q$-exponential function~\cite{app-p2}:
\begin{eqnarray}
\ee_q(x):=[1+(1-q)x]^{\frac{1}{1-q}}~.
\label{qexp}
\end{eqnarray}

Hadron spectra can be described by the Lorentz-invariant particle spectra.
{These were successfully fitted by the non-extensive distributions in a wide center-of-mass energy and $p_T$ range~\cite{all-1, all-2, all-3, all-4, all-5, all-6, all-7, all-8, all-9, all-10,all-10-1,all-11,all-12,all-13}.}
In the following, we focus on the most often used formulas from~\cite{all-1, all-2, all-3, all-4, all-5, all-6, all-7, all-8, all-9} for representing identified particle spectra in various collisions.
This work explores differences between ($m_T-m$) and $m_T$-dependent, as~well as simple $p_T$ functions:
\begin{eqnarray}
E\frac{\dd^3 N}{\dd^3p}=\frac{\dd^3 N}{\dd y p_T\dd p_T \dd\phi}=\frac{1}{2\pi p_T}\frac{\dd^2N}{\dd y\dd p_T} .
\end{eqnarray}

Different research groups used various kinds of expressions of it in order to describe $p_T$ spectra. 
We consider functions of $m_T-m$ and $p_T$ in the non-extensive approach, after applying the normalized functions and the thermodynamically motivated ones~\cite{Shen-2019}.
Our aim is to find the best-fitting functions among these, while assigning a physical interpretation to their parameters. 
We investigate the following distribution forms:
\begin{align}
f_0&=f_{BG}=A_0\cdot \exp\left(\frac{m_T-m}{T_0} \right), \nonumber \\
f_1&=A_1\cdot \left(1+\frac{m_T-m}{n_1T_1}\right)^{-n_1}, \nonumber \\
f_2&=A_2\cdot\frac{(n_2-1)(n_2-2)}{2\pi n_2T_2[n_2T_2+m(n_2-2)]}\cdot\left(1+\frac{m_T-m}{n_2T_2}\right)^{-n_2}, \nonumber \\
f_3&=A_3\cdot m_T\left(1+\frac{m_T-m}{n_3T_3}\right)^{-n_3}, \nonumber \\
f_4&=A_4\cdot \left(1+\frac{m_T}{n_4T_4} \right)^{-n_4}, \nonumber \\
f_5&= A_5\cdot \left(1+\frac{p_T}{n_5T_5}\right)^{-n_5}.
\label{functions}
\end{align} 

{There are relations} among the distributions defined above. 
It is easy to realize that $f_1$ and $f_2$ coincide whenever their amplitudes satisfy the relation
\begin{eqnarray}
A_1=A_2\cdot\frac{(n_2-1)(n_2-2)}{2\pi n_2T_2[n_2T_2+m(n_2-2)]}=A_2\cdot C_q, ~~~\quad {\rm and} \quad~~~n_1=n_2.
\end{eqnarray}

Accounting for the differences between ($m_T-m$) and $m_T$ dependencies, we re-cast $f_1$ and $f_4$ described in Equation (\ref{functions}) as follows:
\begin{eqnarray}
f_1=A_1\cdot \left(1-\frac{m}{n_1T_1}\right)^{-n_1}\cdot \left(1+\frac{m_T}{n_1T_1-m}\right)^{-n_1} .
\end{eqnarray}

Comparing this with $f_4$, we arrive at the relations
\begin{eqnarray}
A_1\cdot \left(1-\frac{m}{n_1T_1}\right)^{-n_1}=A_4,~~~~~
n_1=n_4, \quad {\rm and} \quad~~~n_1T_1-m=n_4T_4.
\label{mmT}
\end{eqnarray}

These comments are important {for the comparison of different approaches}.
They also demonstrate that no inconsistency occurs by applying different fit formulas.
However, differences arise from the statistical physical motivations behind these formulas~\cite{all-1,all-2,all-3,all-4,all-5,all-6,Shen-2019,Shen-2018}.
The corresponding results and discussions are investigated next.
{Note that for all the physical quantities, we use the natural units, $c=1$, for convenience in this paper.}

\section{Results and Discussions}\label{sec:sec2}

In this section, we analyze the transverse momentum distributions of identified pions and kaons stemming from the elementary ($pp$) and heavy-ion ($pPb$ and $PbPb$) collisions fitted by the functions listed in Equation (\ref{functions}).
All the relevant parameters are then analyzed in order to investigate further the non-extensive physics behind these collisions. 

\subsection{Analysis of the $pp$ Spectra}

\begin{figure}[!htb]
\centering
\scalebox{1}[1]{
\includegraphics[width=0.48\linewidth]{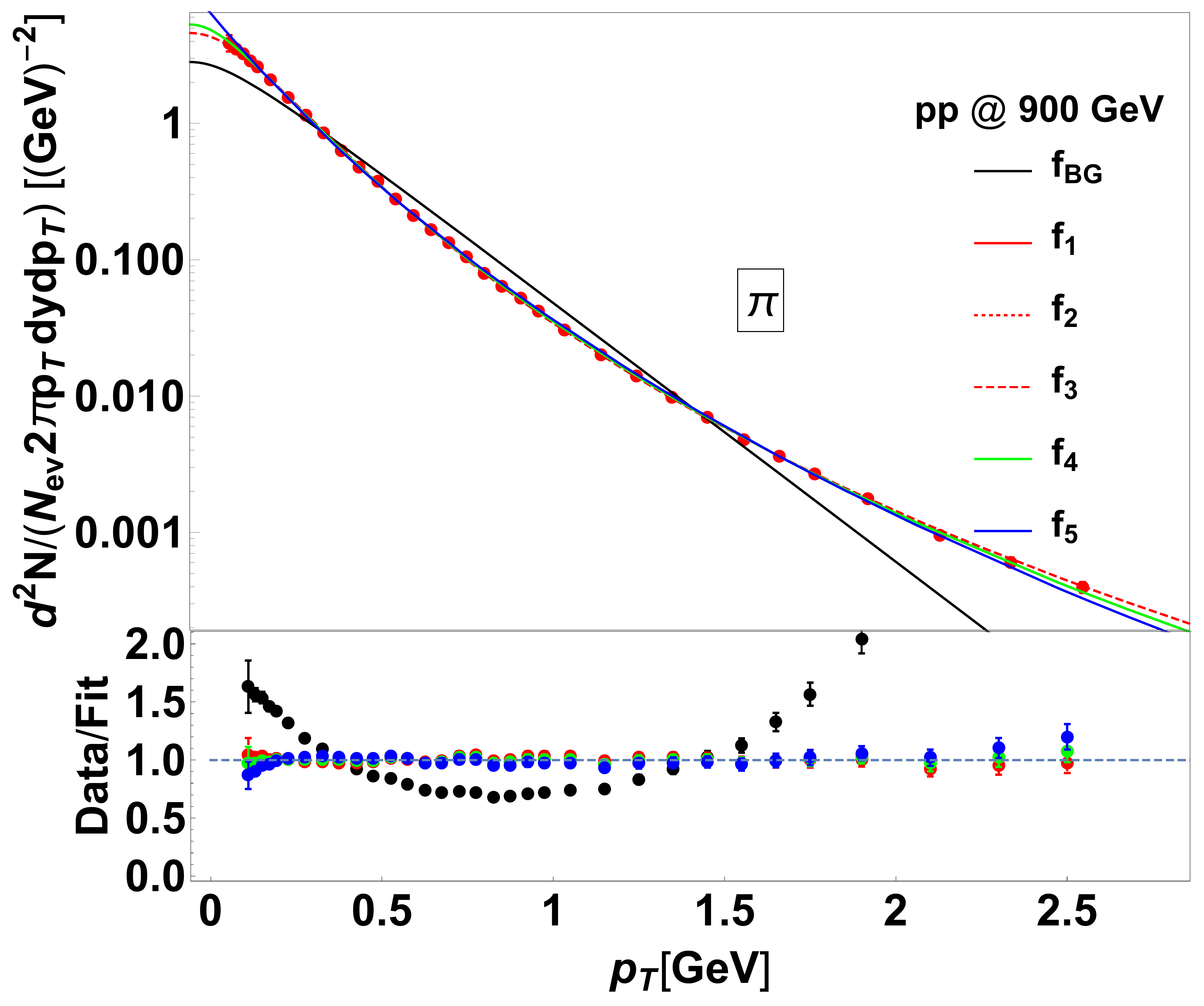}
\includegraphics[width=0.48\linewidth]{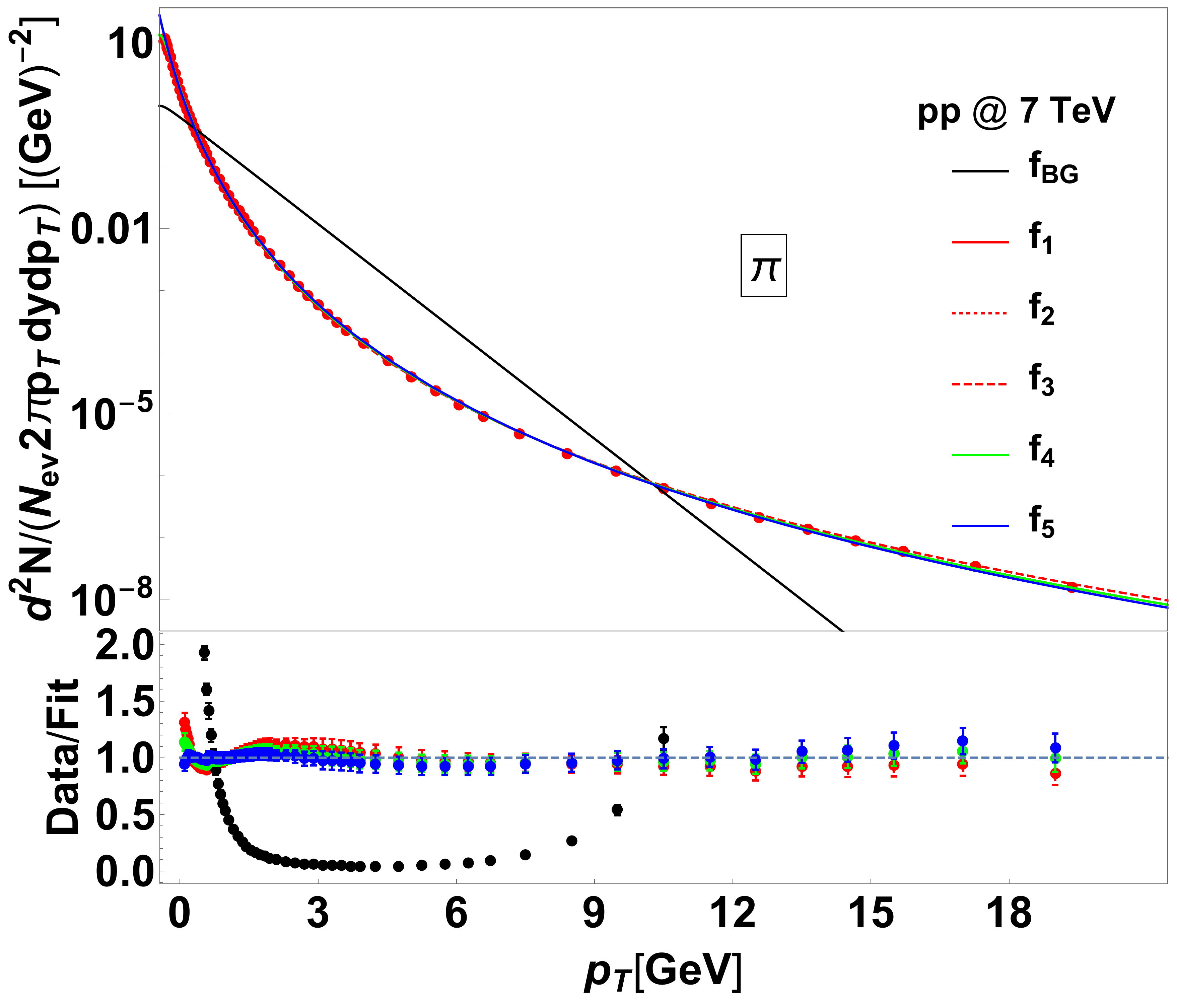}
}
\vspace{-5mm}
\scalebox{1}[1]{
\includegraphics[width=0.48\linewidth]{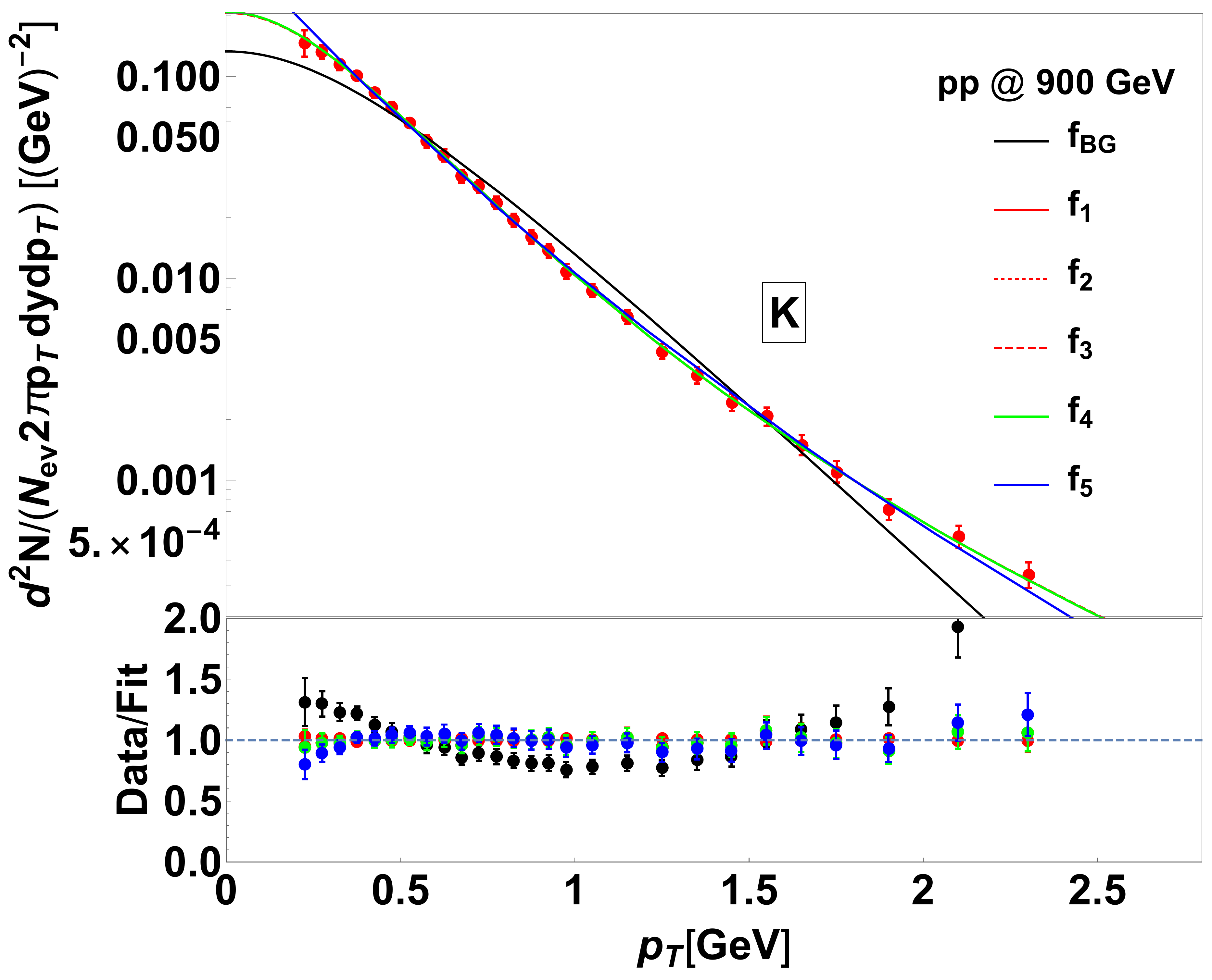}
\includegraphics[width=0.48\linewidth]{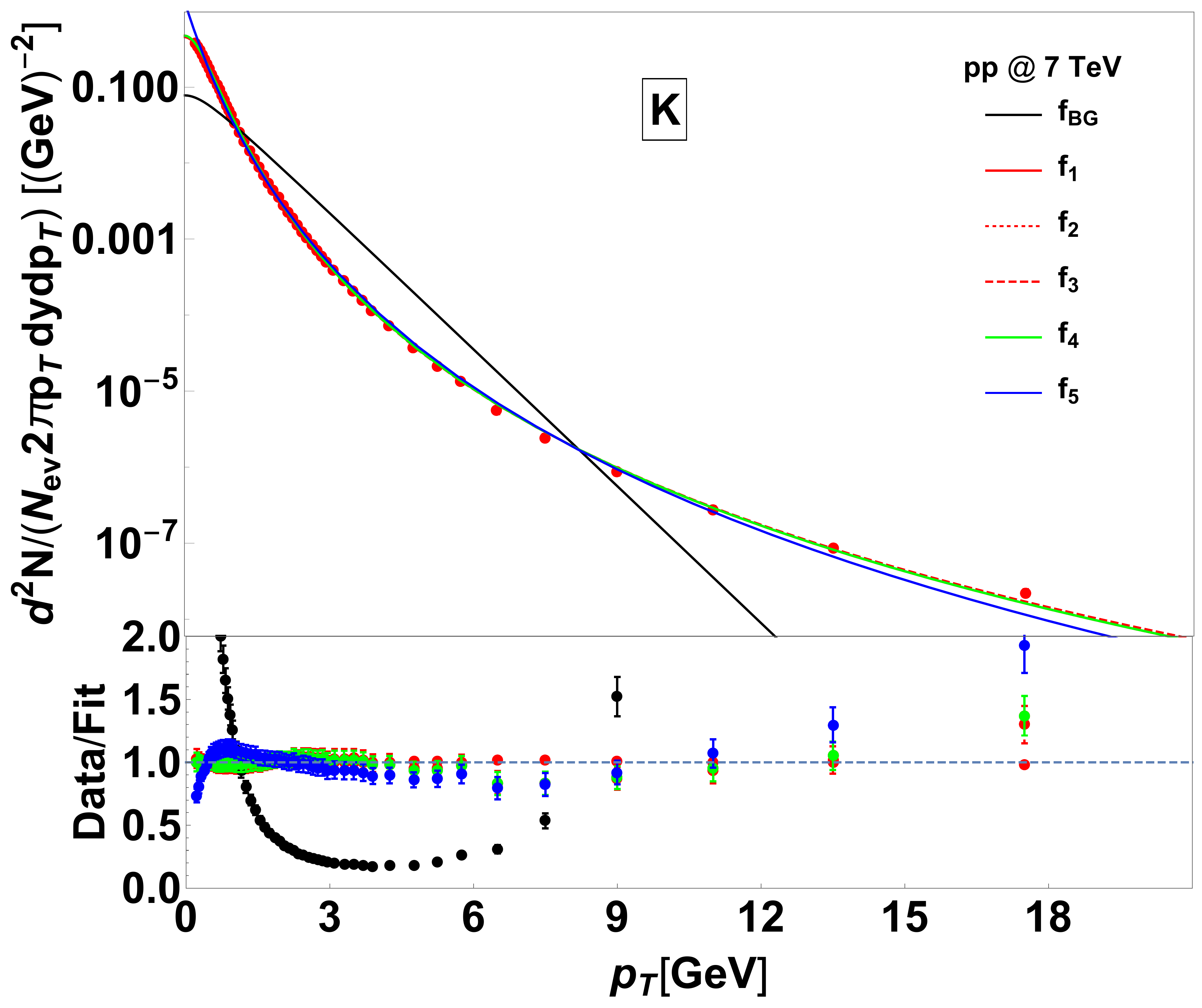}
}
\caption{The $p_T$ spectra for pions (upper) and kaons (lower) in $pp$ collisions at $\sqrt{s}=900$ GeV and 7 TeV at midrapidity as examples. Data are taken from Refs.\cite{all-10,all-10-1}. All are fitted with all the functions of Eq.(\ref{functions}) in the ranges of $0.1<p_T<2.6$ GeV at $\sqrt{s}=$900~GeV and $0.1<p_T<20$ GeV at 7~TeV, respectively. Ratios of the net fits to data are also shown in the lower panel. The relevant values of $\chi^2/d.o.f.$ are shown in Table \ref{tabchi}.}
\label{figppk}
\end{figure}

In high-energy physics, even the smallest hadron--hadron ($pp$) collisions are rather complicated processes.
One usually separates two main regimes of hadron production:
one is a soft multiparticle production, dominant at low transverse momenta, where the spectra can also be fitted by an exponential behavior~\cite{Hagedorn-1995}, cf. the curve $f_{BG}$ in Figure \ref{figppk}.
We realize that $f_{BG}$  describes well this part of the spectra even in $pp$ collisions.
As $p_T$ gets higher ($p_T>$3 {GeV}), the spectrum displays a power-law tail.
They~are predicted by perturbative QCD, 
 owing to the hard scattering of current quarks and gluons.
In a number of publications \cite{all-10,all-10-1,all-11,all-12,all-13}, the Tsallis statistical distribution was successfully applied to describe data for $pp$ collisions over a wide range of the transverse momenta because of its two limits: the exponential shape at small $p_T$ and the power-like distribution at large $p_T$,
\begin{eqnarray}
\ee_q(-\frac{p_T}{T}) \longrightarrow \left\{ \begin{array}{l}  
		~\ee^{-p_T/T} \qquad  ~~~~~~~~~~~~~~ p_T\to 0 \\ 
		\\	
		\Big( (q-1)\frac{p_T}{T}\Big)^{\frac{1}{1-q}}  \qquad  p_T\to \infty 	.
\end{array} \right. ~.
\end{eqnarray}
\unskip
\unskip
\begin{table}[!htb]
\caption{The values of $\chi^2/d.o.f.$ of spectral fits for pions, kaons, and~protons in $pp$ collisions at 900~GeV and 7~TeV as examples.}
\centering
\begin{tabular}{cccccccc}
\toprule
\textbf{Collision Energy (\boldmath{$\sqrt{s}$})} & \textbf{Produced Hadrons} & \boldmath{$f_{BG}$} & \boldmath{$f_1$} & \boldmath{$f_2$} & \boldmath{$f_3$} & \boldmath{$f_4$} & \boldmath{$f_5$} \\
\hline
\hline
~~~~~~~~~~~~~~~& $\pi$ & 110.8 & 0.2814 & 0.2814 & 0.4697 & 0.2814  &  1.456 \\
900~GeV & $K$  &  8.047 & 0.1748 & 0.1749 & 0.1698 & 0.1749 & 0.6669 \\ 
~~~~~~~~~~~~~~~& $p$  &  3.491 & 0.3724 & 0.3724 & 0.3735 & 0.3724 & 0.4145 \\
\hline
~~~~~~~~~~~~~~& $\pi$ & 1316.0 & 0.9681 & 0.9681 & 3.417 & 0.9681 & 0.3049 \\ 
7~TeV~~~~& $K$   &  520.2 & 0.4202 & 0.4202 & 0.4313 & 0.4202 & 3.100 \\
~~~~~~~~~~~~~~~& $p$  & 254.3 & 0.4481 & 0.4481 & 0.4356 & 0.4481 & 4.357 \\ 
\hline
\hline
\end{tabular}
\label{tabchi}
\end{table}

We focus on the fittings of the produced charged particle spectra in elementary collisions with the non-extensive functions in Equation (\ref{functions}).
Data were taken for pions, kaons, and~protons in $pp$ collisions at $\sqrt{s}=62.4$~GeV, $200$~GeV {from the PHENIX Collaboration}~\cite{all-11} and at 900~GeV~\cite{all-10}, 2.76~TeV~\cite{all-12}, 5.02~TeV, and~7~TeV~\cite{all-10-1} {from the ALICE Collaboration}. 
We restrict our analysis to the midrapidity region $|y|<0.5$ within the $p_T$ ranges, as shown in {Table \ref{tabpp}}. 
Note that in the following, $\pi$, $K$, and~$p$ mark the spectra of $\frac{\pi^++\pi^-}{2}$, $\frac{K^++K^-}{2}$, and~$\frac{p+\bar{p}}{2}$, respectively.
\begin{table}[!htb]
\caption{Fitting $p_T$ ranges of spectra for different charged particles in $pp$ collisions~\cite{all-10,all-10-1,all-11,all-12}.}
\centering
\begin{tabular}{cccc}
\toprule
\boldmath{$\sqrt{s}$} & \boldmath{$\pi$} \textbf{[GeV]} & \boldmath{$K$} \textbf{[GeV] }& \boldmath{$p$} \textbf{[GeV]} \\
\hline
\hline
62.4~GeV & 0.3--2.9 
& 0.4--2 & 0.6--3.6 \\
200~GeV  &  0.3--3  & 0.4--2 &  0.5--4.6 \\ 
900~GeV  & 0.1--2.6 & 0.2--2.4 & 0.35--2.4 \\
2.76~TeV  & 0.1--20 & 0.2--20 & 0.3--20 \\ 
5.02~TeV   &  0.1--20  & 0.2--20   &  0.3--20 \\
7~TeV  & 0.1--20 & 0.2--20 & 0.3--20 \\ 
\hline
\hline
\end{tabular}
\label{tabpp}
\end{table}

Figure \ref{figppk} shows that all of the different non-extensive functions we used fit the pion and kaon spectra very well for various kinds of beam energies at midrapidity.  
{The ratios of $\chi^2/d.o.f.$ of the relevant fits are given in Table \ref{tabchi}.} 
Specifically, the first two distributions ($f_1$ and $f_2$) of $m_T-m$ and $f_4$ of $m_T$ show close-fitting results.
The distribution, $f_3$, derived thermodynamically, does not display large differences in the goodness of fit either.
Checking the fitting parameters $A$, $T$, and~$q=1+1/n$, we~observe that, as we expected and introduced in the previous section, all these functions share the same Tsallis parameter $n$.
The two $m_T-m$ functions ($f_1$ and $f_2$)
{lead to fitting values of the temperature $T$, which are different from the pure $m_T$ fit ($f_4$).}
This indicates that the normalization constant does not affect the fitted $T$ and $q$ parameters but the integrated yield $\dd N/\dd y$.
Namely, by normalizing the momentum spectrum
\begin{eqnarray}
\frac{1}{2\pi p_T}\frac{\dd^2N}{\dd y \dd p_T}=A_2\cdot C_q\cdot \left(1+\frac{m_T-m}{n_2T_2}\right)^{-n_2}
\end{eqnarray}
with the $C_q$ normalization constant and the condition of $A_2=\dd N/\dd y$, we obtain the integral over $p_T$ from 0 to its {maximal} values $p_{Tmax}$:
\begin{eqnarray}
\int_0^{p_{Tmax}} \frac{1}{2\pi p_T}\frac{\dd^2N}{\dd y \dd p_T}2\pi p_T \dd p_T =\frac{\dd N}{\dd y}~{.}
\end{eqnarray}

Moving towards physical interpretation issues, we investigate the temperature, $T$, and~the non-extensive parameter, $q$.
Investigations in~\cite{all-11,Cleymans-2013} showed that both of them express $\sqrt{s}$ dependence.
In this paper, we found that they are also dependent on the hadron mass, $m$.
The $\sqrt{s}/m$ dependence, as a result, is studied in order to analyze hadron spectra parameters within the non-extensive approach.
{Following the phenomenological observations in~\cite{Gergely-1,Gergely-2}, a QCD-like evolution can be introduced for both the parameters $T$ and $q$. 
While analyzing data, we found that the temperature $T$ had a weak logarithmic $\sqrt{s}/m$ dependence. Thus, here we assume a linear $\sqrt{s}/m$ dependence to analyze the temperature $T$, but the non-extensive parameter $q$ is kept with the stronger logarithmic distribution: }
\begin{eqnarray}
T=T_0+T_1\cdot\Big(\frac{\sqrt{s}}{m}\Big)~, \quad {\rm and} \quad~ q=q_0+q_1\cdot\ln\Big(\frac{\sqrt{s}}{m}\Big)~.
\label{Tqsm}
\end{eqnarray}

\begin{figure}[!htb]
\centering
\includegraphics[width=0.42\linewidth]{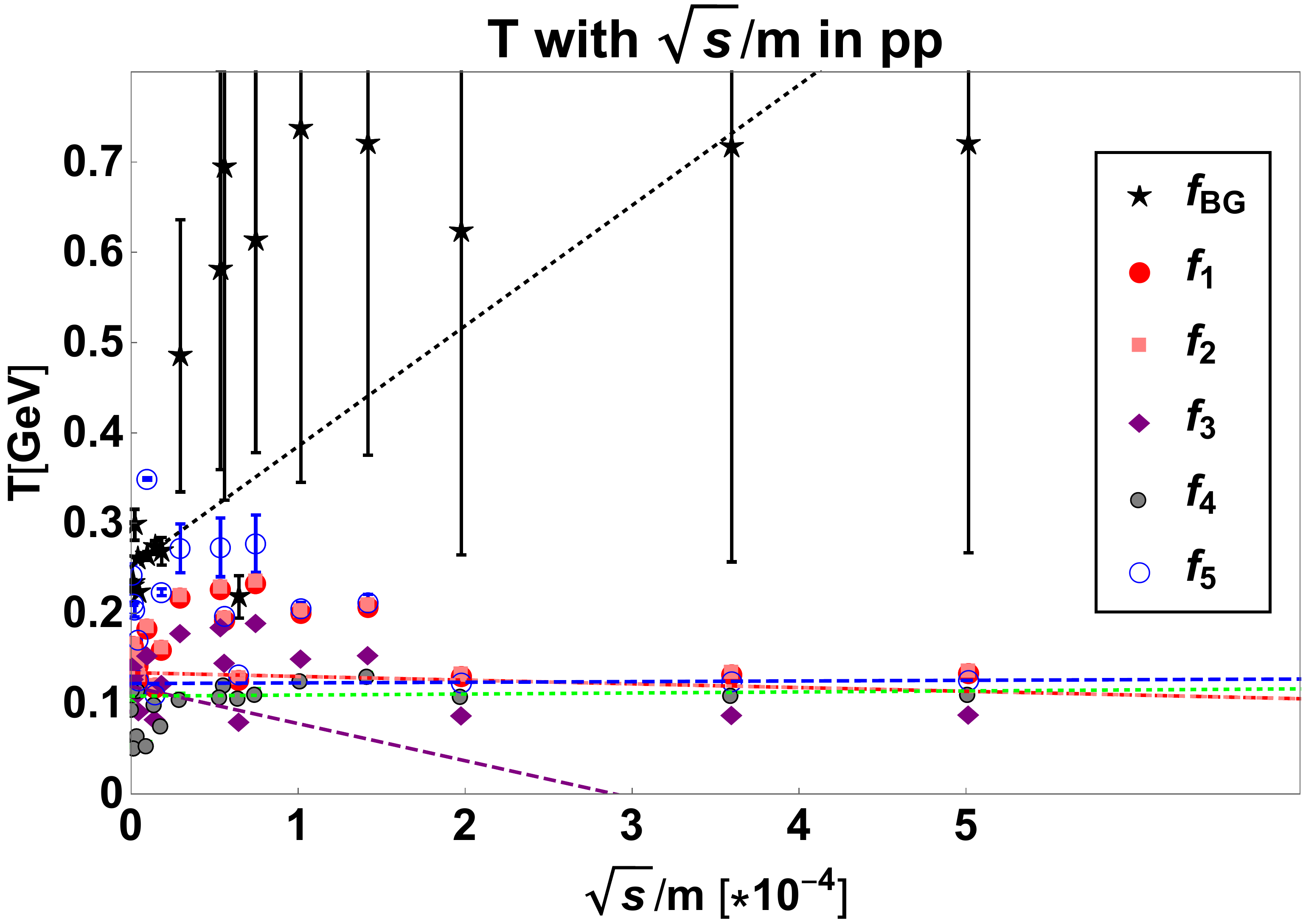}
\includegraphics[width=0.42\linewidth]{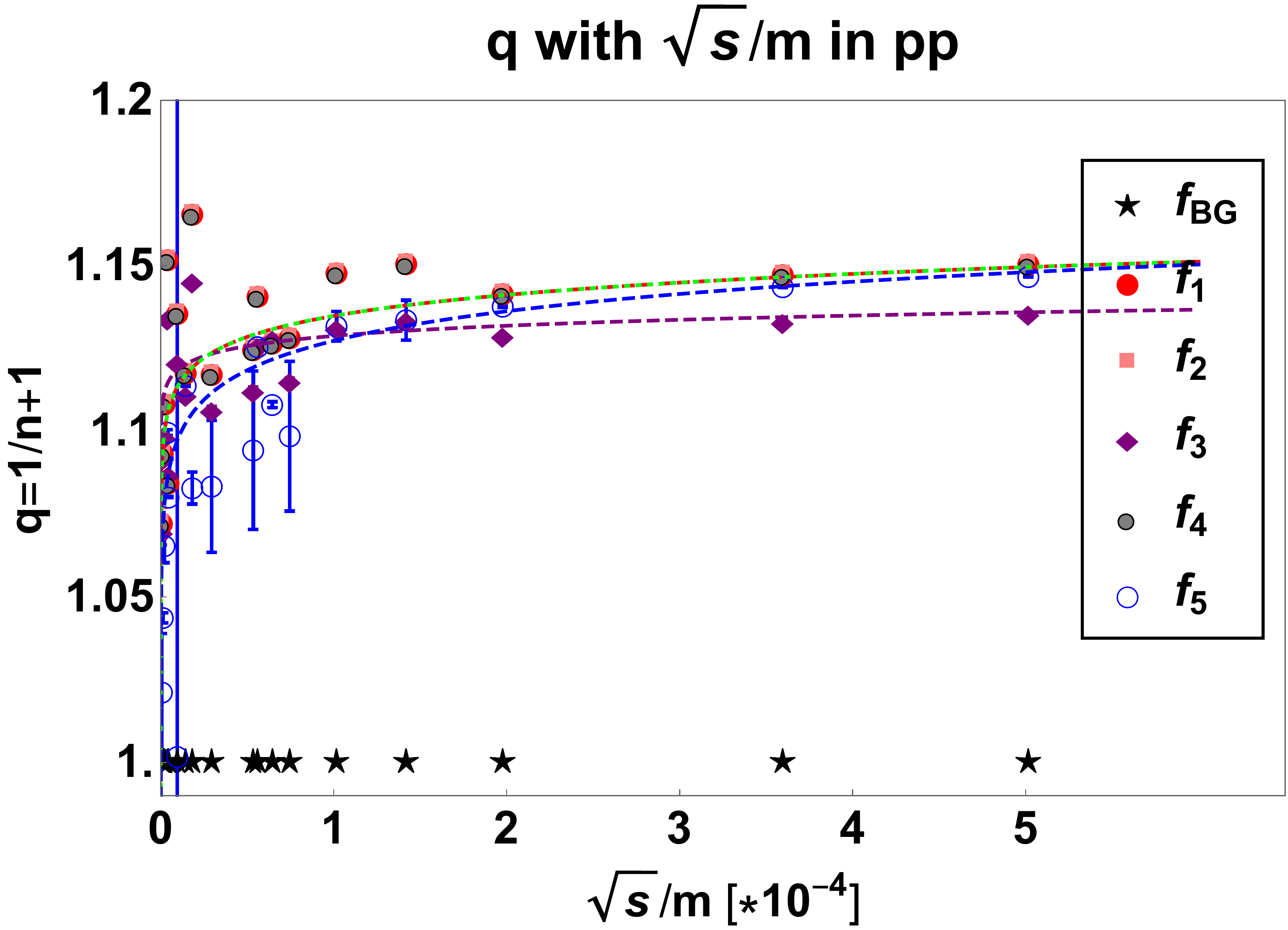}
\caption{Both the center-of-mass energy $\sqrt{s}$ and hadron mass $m$ distributions of the fitting temperature $T$ and the non-extensive parameter $q$. In this work, we analyze the results at all given energies with the relationship cf. Equation (\ref{Tqsm}). 
	Here we list the results for $\sqrt{s}=$62.4~GeV, 200~GeV, 900~GeV, 2.76~TeV, 5.02~TeV, and~7~TeV. and hadron species of pions, kaons, and~protons. We have extracted a factor of $10^4$ from the values of $\sqrt{s}/m$ for convenience.}
\label{figTq}
\end{figure}

In summary, our work indicates that the BG distribution is not suitable for describing the hadron spectra over a wide range of $p_T$.
Comparisons of their corresponding fitting errors {$\chi^2/d.o.f.$} show that both $m_T-m$ and $m_T$ functions share the same goodness between $f_1$ and $f_2$, cf. Equation (\ref{functions}).
Together with the {thermodynamically derived} $f_3$, all the non-extensive approaches ($f_1\sim f_4$) follow the experimental data accurately. 
The fitting temperature, $T$, is nearly {constant when changing} the ratio of the collision energy to hadron mass, $\sqrt{s}/m$.
Specifically, distributions of $f_1$, $f_2$, $f_4$, and~$f_5$ are described best with such a connection, as shown in the left panel of Figure \ref{figTq}.
From Table \ref{tabTqsm}, {we also see that the slope} parameters in these four cases are almost zero, which means that they are constant around some values.
The non-extensive parameter $q$, on the other hand, follows a logarithmic dependence, agreeing with a pQCD-based motivation, cf.~\cite{Shen-2019}.
Note that our results on $T$ and $q$ are different from the work by Cleymans et al.~\cite{Cleymans-2013}.
Those authors parameterized this relation as a power-law.
\begin{table}[!htb]
\caption{Fitting parameters of Equation (\ref{Tqsm}) in use within Figure \ref{figTq}:}
\centering
\begin{tabular}{ccccc}
\toprule
{\bf Fitting Functions} & \boldmath{$T_0$} & \boldmath{$T_1$} & \boldmath{$q_0$} & \boldmath{$q_1$} \\
\hline
$f_{BG}$ & 0.2515~$\pm$~0.0005 & 0.1335~$\pm$~0.0002 & - & - \\
$f_1$  & 0.1343~$\pm$~0.0003 & $-$0.0041~$\pm$~0.0001   & 1.135~$\pm$~0.002 & 0.009~$\pm$~0.001 \\ 
$f_2$  & 0.1343~$\pm$~0.0003 & $-$0.0041~$\pm$~0.0001 & 1.135~$\pm$~0.002 & 0.009~$\pm$~0.001 \\
$f_3$  & 0.1190~$\pm$~0.0002 & $-$0.0412~$\pm$~0.0002 &  1.129~$\pm$~0.001  & 0.004~$\pm$~0.001 \\ 
$f_4$  & 0.1083~$\pm$~0.0003  & 0.0011~$\pm$~0.0004   &  1.135~$\pm$~0.002 & 0.009~$\pm$~0.001\\
$f_5$  & 0.1222~$\pm$~0.0005  & 0.0007~$\pm$~0.0001 & 1.127~$\pm$~0.002  & 0.013~$\pm$~0.002 \\ 
\hline
\end{tabular}
\label{tabTqsm}
\end{table}

\subsection{Analysis of the $pPb$ and $PbPb$ Results}\label{sec:sec4}

In $pPb$~\cite{all-10-1} collisions at 5.02~TeV and in $PbPb$~\cite{AA-1,AA-2,AA-3,AA-4} collisions at 2.76~TeV, more kinds of hadron spectra are analyzed within the formulas of Equation (\ref{functions}). 
Data are taken from the ALICE Collaboration within wide $p_T$ ranges, as seen in Table \ref{tabNN}. 
We observe that all of them present good fittings over the whole range of $p_T$ for each hadron at various kinds of centrality bins.
On the other hand, similar to the $pp$ cases, the BG formula can still perform well just in the low $p_T$ region ($p_T<3$ {GeV}).

\begin{table}[!htb]
\caption{Fitting $p_T$ range of different hadron spectra in heavy-ion collisions in this work~\cite{all-10-1,AA-1,AA-2,AA-3,AA-4}.}
\centering
\begin{tabular}{cccc}
\toprule
{\bf Particles} &  {\bf Mass [GeV]} & {\bf \boldmath{$pPb$} [GeV]}  & {\bf \boldmath{$PbPb$} [GeV]}  \\
\hline
$\pi$ & 0.140 & 0.11--2.85 
& 0.11--19 \\
$K$ & 0.494 &  0.225--2.45 & 0.225--19 \\
$K_S^0$ & 0.498 & 0.05--7 &  0.45--11 \\
$K^*$ &  0.896 &    &  0.55--4.5 \\
$p$ & 0.938 &  0.325--3.9  & 0.325--17.5 \\
$\phi$ &  1.019 &    &  0.65--4.5 \\
$\Lambda$ & 1.116 &  0.65--7 & 0.65--11 \\
$\Xi$ &  1.321 &    &  0.7--7.5 \\
$\Omega$ &  1.672 &    &  1.3--7.5 \\
\hline
\hline
\end{tabular}
\label{tabNN}
\end{table}

\begin{figure}[!htb]
\centering
\scalebox{1}[1]{
\includegraphics[width=0.45\linewidth]{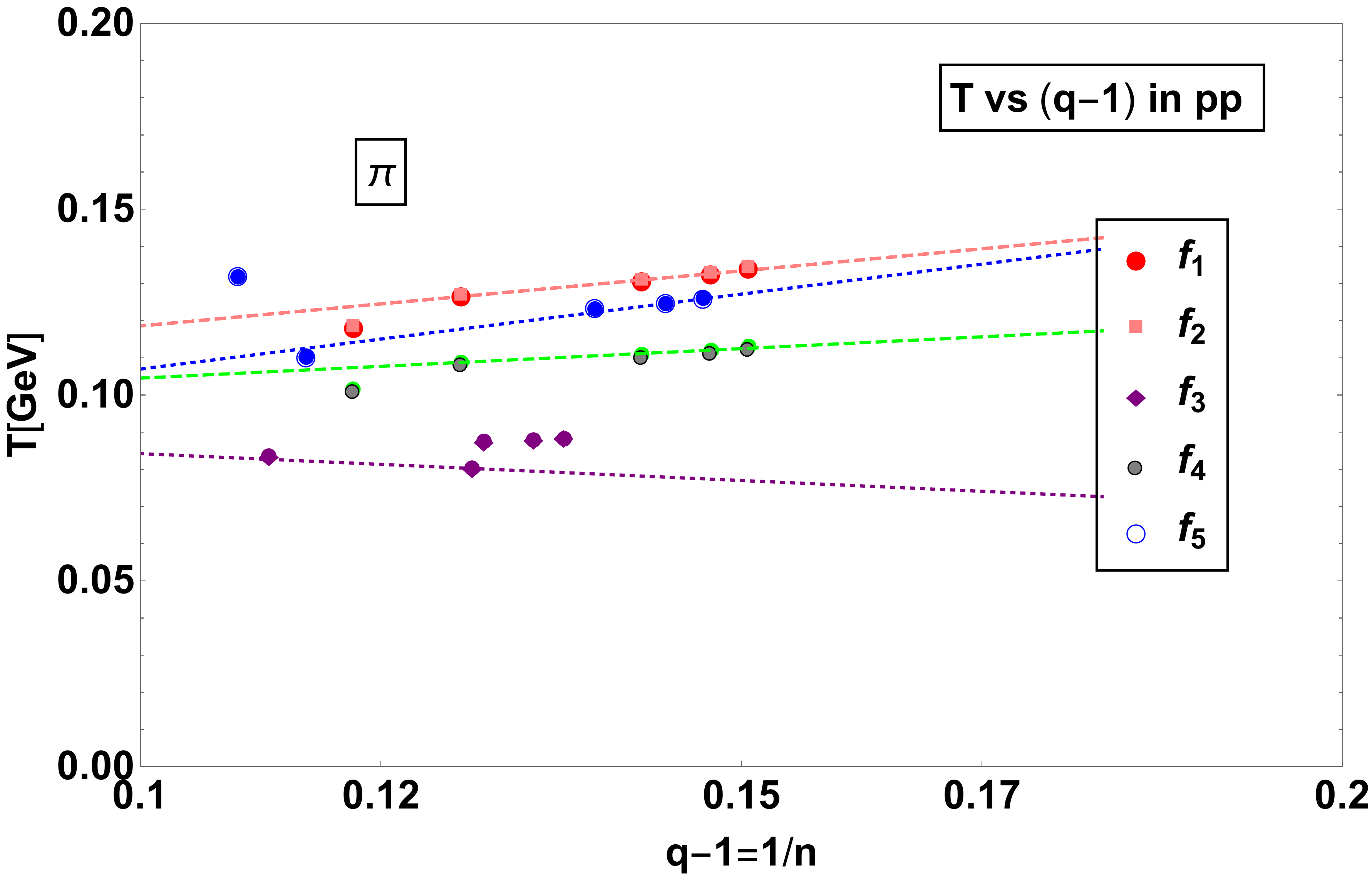}
\includegraphics[width=0.45\linewidth]{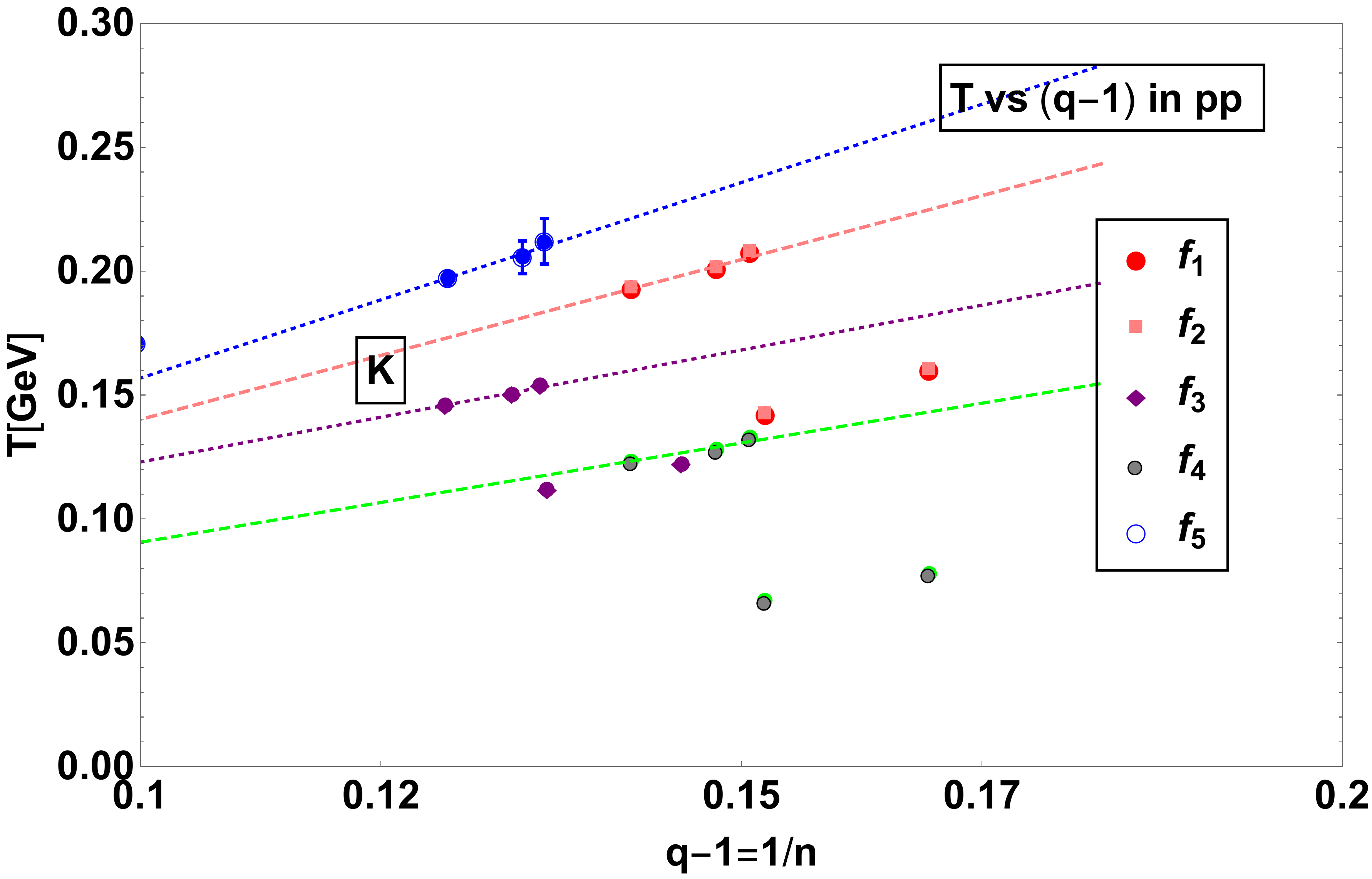}
}
\scalebox{1}[1]{
\includegraphics[width=0.45\linewidth]{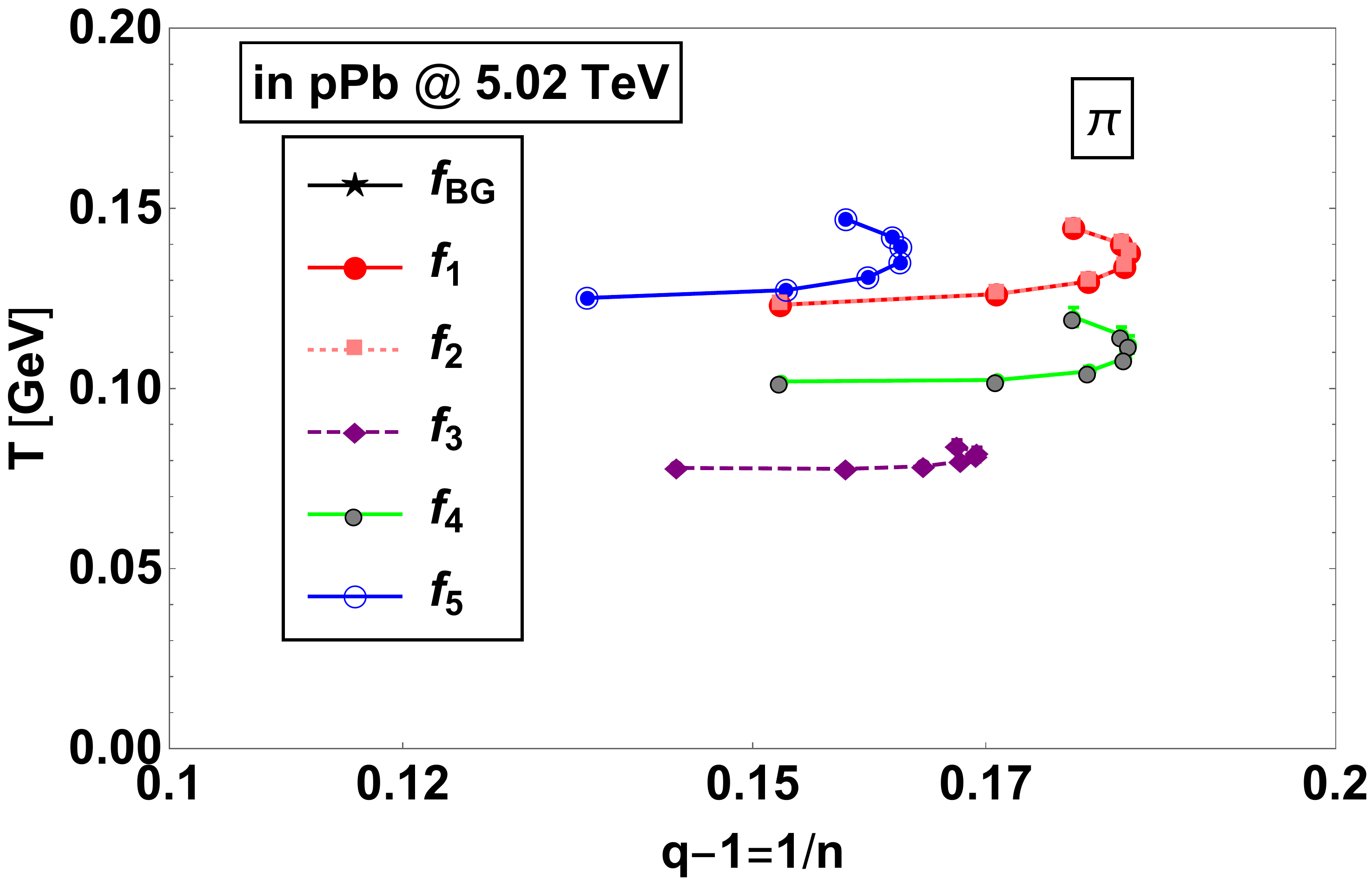}
\includegraphics[width=0.45\linewidth]{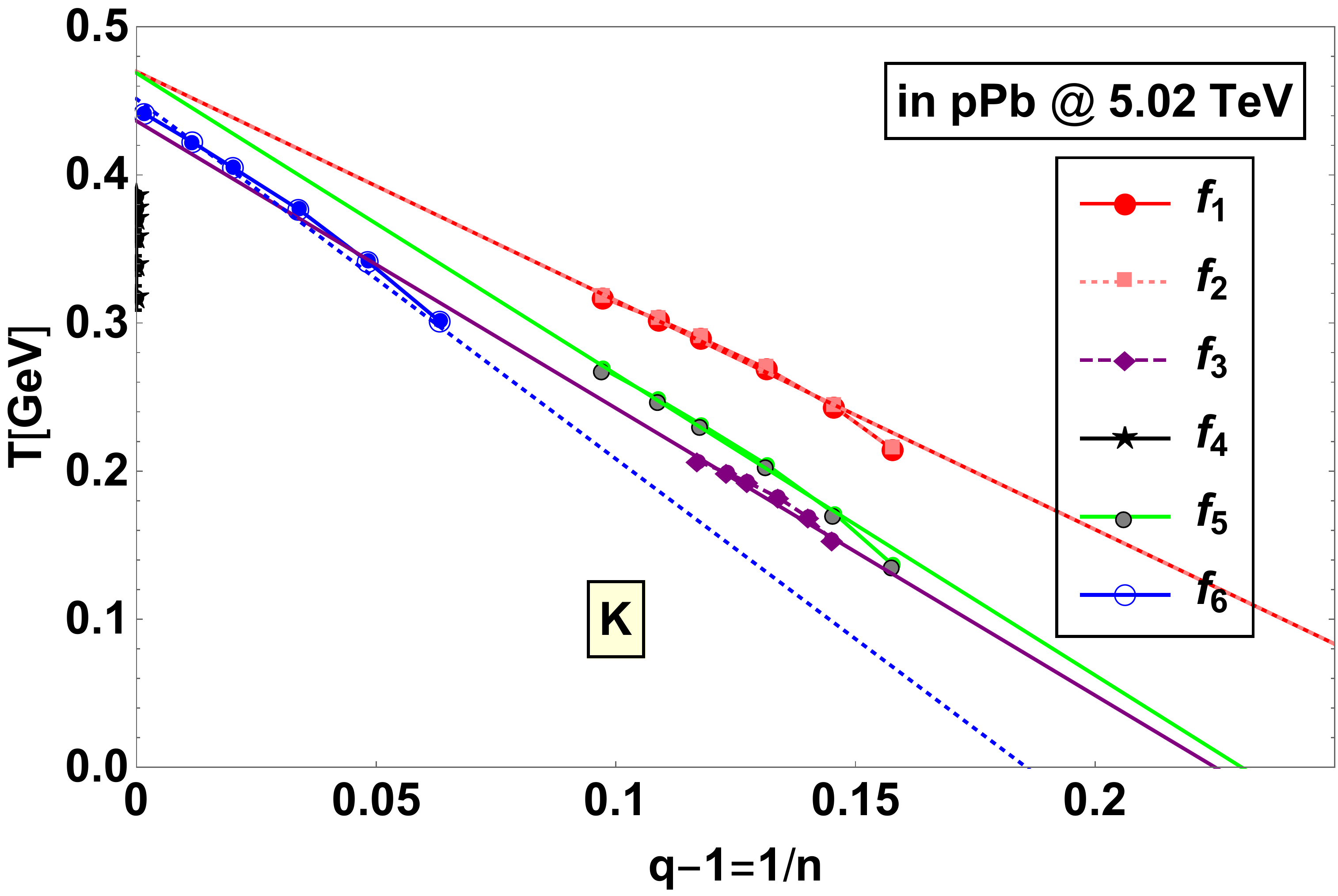}
}
\scalebox{1}[1]{
\includegraphics[width=0.45\linewidth]{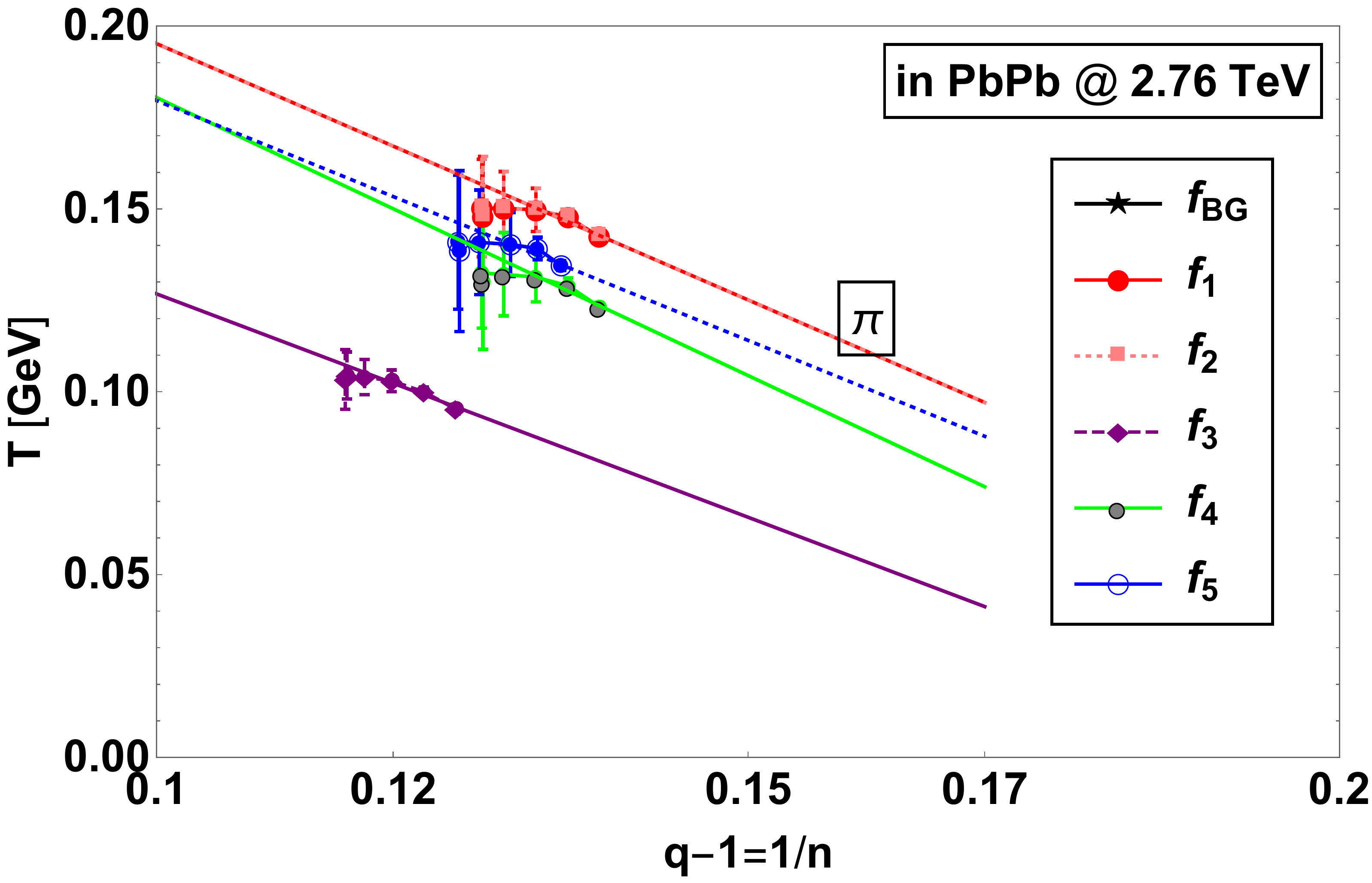}
\includegraphics[width=0.45\linewidth]{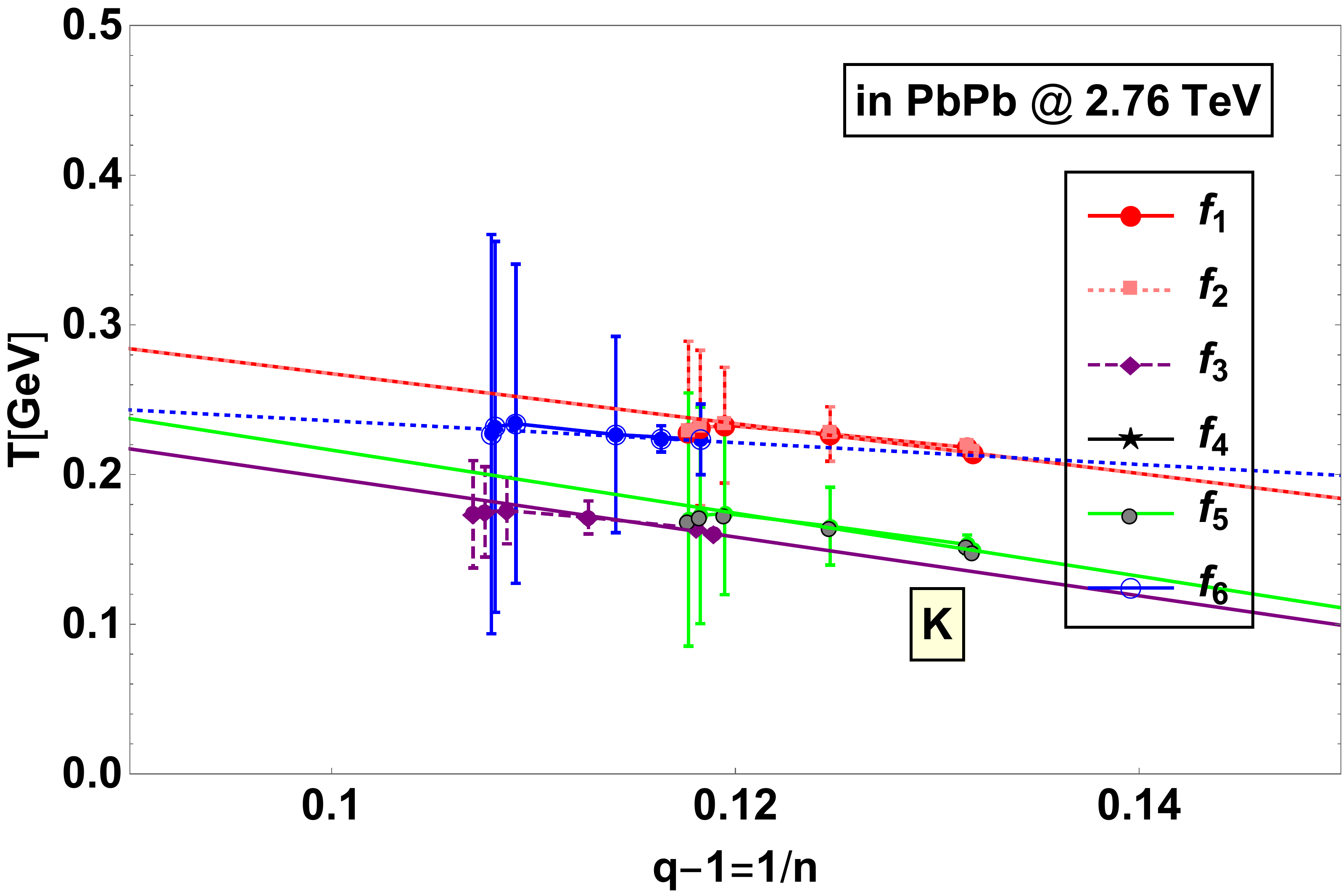}
}
\caption{Correlations between $T$ and $q-1=1/n$ for spectra of $\pi$ (left) and $K$ (right) in $pp$, $pPb$, and~$PbPb$ collisions. The corresponding $p_T$ range is listed in Table \ref{tabNN}, and~the values of fitting parameters in Equation (\ref{Tqrel}) are listed in Table \ref{tabTqNN}. }
\label{figklTq}
\end{figure}

In this work, as an example, we analyzed the fitting results of $p_T$ spectra of pions and kaons produced in all kinds of collisions mentioned above.
It is instructive to plot the relationship between the fitting temperature $T$ and the Tsallis parameter $q$ for the same hadron spectra for different centralities in the same heavy-ion collisions.
The results of pions and kaons in $pp$ collisions are also analyzed as comparisons. 
In Figure \ref{figklTq}, we show the linear correlating appearances for both $\pi$ and $K$ in $pPb$ at 2.76~TeV~\cite{all-10-1} and in $PbPb$ at 5.02~TeV~\cite{AA-1, AA-2} as well as the $pp$ results in all kinds of collision energies~\cite{all-10,all-10-1,all-11,all-12} in this paper.
In fact, whatever kinds of particle we study, all these non-extensive fittings {give a similar dependence} of $T$ on the parameter $q$:
\begin{eqnarray}
T\approx T_0-(q-1)T_1~,
\label{Tqrel}
\end{eqnarray} 
which agrees with our previous work~\cite{Shen-2018,Shen-2019} and that of others~\cite{Wilk-2015}.

Note that the slope parameter $T_1$ in Table \ref{tabTqNN} turns negative and $T_0$ is nearly zero for the $pp$ case, as discussed in~\cite{Shen-2018}.
Results of fittings on pion spectra, typically in $pPb$ collisions at 5.02~TeV, fail in the obvious linear combinations probably due to the small mass of pions and high multiplicities.
It is found that all forms of non-extensive distributions feature a similar relation between the temperature $T$ and non-extensive parameter $q$.
This, in turn, hopefully promotes a better understanding of the meaning of the non-extensive parameter $q$.

\begin{table}[!htb]
\caption{Fitting parameters of Equation (\ref{Tqrel}) between $T$ and $q-1=1/n$ for spectra of $\pi$ (upper) and $K$ (lower) in $pp$, $pPb$, and~$PbPb$ collisions (note that $f_{BG}$ is not included because $q=1$ is a constant).}
\centering
\scalebox{0.92}[0.92]{
\begin{tabular}{cccccccc}
\toprule
{\bf Particles} & {\bf Fittings} & \boldmath{$T_1$ \textbf{in} $pp$} & \boldmath{$T_0$ \textbf{in} $pp$} & \boldmath{$T_1$ \textbf{in} $pPb$} & \boldmath{$T_0$ \textbf{in} $pPb$} & \boldmath{$T_1$ \textbf{in} $PbPb$} & \boldmath{$T_0$ \textbf{in} $PbPb$} \\
\hline
\hline
~ & $f_1$  & $-$0.36~$\pm$~0.02 & 0.08~$\pm$~0.01  & - & - & 1.40~$\pm$~0.02 & 0.335~$\pm$~0.004 \\ 
~ & $f_2$  & $-$0.36~$\pm$~0.02 & 0.08~$\pm$~0.01 & - & -  & 1.40~$\pm$~0.02 & 0.335~$\pm$~0.004 \\
$\pi$ & $f_3$  & $-$0.14~$\pm$~0.04 & 0.07~$\pm$~0.02 & - & - &  1.22~$\pm$~0.02  & 0.249~$\pm$~0.005 \\ 
~ & $f_4$  & $-$0.22~$\pm$~0.01  & 0.08~$\pm$~0.01   & - & - &  1.52~$\pm$~0.02 & 0.333~$\pm$~0.005 \\
~ & $f_5$  & $-$0.31~$\pm$~0.03  & 0.08~$\pm$~0.01 & - & -  & 1.31~$\pm$~0.01  & 0.311~$\pm$~0.007 \\ 
\hline
~ & $f_1$  & $-$1.30~$\pm$~0.02 & 0.011~$\pm$~0.001   & 1.55~$\pm$~0.02 & 0.470~$\pm$~0.001 & 1.67~$\pm$~0.06 & 0.434~$\pm$~0.003\\ 
~ & $f_2$  & $-$1.30~$\pm$~0.02 & 0.011~$\pm$~0.001 & 1.55~$\pm$~0.02 & 0.470~$\pm$~0.001 & 1.67~$\pm$~0.06 & 0.434~$\pm$~0.003 \\
$K$ & $f_3$  & $-$0.90~$\pm$~0.04 & 0.032~$\pm$~0.005 &  1.94~$\pm$~0.03  & 0.436~$\pm$~0.004 & 1.96~$\pm$~0.03 & 0.394~$\pm$~0.007 \\ 
~ & $f_4$  & $-$0.81~$\pm$~0.01  & 0.010~$\pm$~0.004   &  2.03~$\pm$~0.03 & 0.470~$\pm$~0.003 & 2.11~$\pm$~0.05 & 0.427~$\pm$~0.006 \\
~ & $f_5$  & $-$1.59~$\pm$~0.02  & $-$0.001~$\pm$~0.0005 & 2.43~$\pm$~0.01  & 0.453~$\pm$~0.002 & 0.73~$\pm$~0.02 & 0.309~$\pm$~0.007 \\ 
\hline
\hline
\end{tabular}}
\label{tabTqNN}
\end{table}

\section{Summary}\label{sec:sec3}

In this work, we analyzed various {fitting formulas of the hadron spectra} in order to explore their sensitivity to different fitting parameters in use within the non-extensive approaches, cf. Equation~(\ref{functions}). 
The hadronization, as well as the $p_T$ distributions in high-energy physics (in proton--proton, {proton--nucleus}, and~nucleus--nucleus collisions) are being studied here. For more details, see~\cite{Shen-2019}.  

Our results reveal that {normalization parameters have} no major effect on the shape of these functions.
In other words, the fitting formulas of either $m_T-m$ or $m_T$ lead to the same fit quality.
As shown in Table \ref{tabchi}, they obtained similar fitting values of $\chi^2/d.o.f.$
Finally, we {investigated} the relationship between the fitting parameters, $T$ and $q$. 
{In $pp$ collisions, the temperature values were fitted by the linear relation of $\sqrt{s}/m$, while the non-extensive parameter $q$ had a logarithmic $\sqrt{s}/m$ dependence, motivated by the QCD-like evolution~\cite{Gergely-1,Gergely-2}.}~All kinds of approaches led to linear relations between the temperature, $T$, and~the non-extensive parameter, $q-1$, in heavy-ion collisions at different centralities. 
This agrees well with our previous results~\cite{Shen-2018, Shen-2019} and others in~\cite{Wilk-2015}. 

Summarizing, based on the Tsallis $q$-exponential, five types of non-extensive formulas in Equation~(\ref{functions}) were investigated in parallel to the usual BG distribution.
Results showed that the BG statistics failed in describing the hadronization in the whole $p_T$ range. 
Within the non-extensive approaches, $m_T-m$ functions obtained similar fitting results to the $m_T$ ones.
This provides a free choice between the functions $m_T-m$ and $m_T$ when analyzing the hadron spectra. 
On the other hand, it does not make any differences with regards to the normalization.
Nevertheless, the normalized function, $f_2$, is the best choice since it is also connected to the particle yield per unit rapidity, $\dd N/\dd y$, by its normalization, $A_2$.

\vspace{0.3cm}

\noindent {\bf Acknowledgments}

\vspace{3mm}

This work has been supported by the Hungarian National Research, Development and Innovation Office (NKFIH) under the contract numbers K120660 and K123815 and THOR COST CA 15213.

\end{document}